\documentclass[12pt]{article}

\title{The more precise determination of hadronic contribution to muonic
$(g-2)$ factor and to $\alpha(M^2_z)$.}
\author{B.V.Geshkenbein
\thanks{E-mail address: geshken@heron.itep.ru}\\ Institute of Theoretical
and Experimental Physics,\\
B.Cheremushkinskaya 25, 117218 Moscow,Russia}
\date{}
\begin{document}
\maketitle

\newcommand{\be}{\begin{equation}}
\newcommand{\ee}{\end{equation}}

\def\la{\mathrel{\mathpalette\fun <}}
\def\ga{\mathrel{\mathpalette\fun >}}
\def\fun#1#2{\lower3.6pt\vbox{\baselineskip0pt\lineskip.9pt
\ialign{$\mathsurround=0pt#1\hfil##\hfil$\crcr#2\crcr\sim\crcr}}}

\begin{abstract}

The hadronic vacuum-polarization contribution to the muon anomalous magnetic
moment $a_{\mu}(hadr.)$ and to the value of the eleectromagnetic coupling
constant at $q^2 = M^2_z$ are determined more precisely in connection with a
new more exact value of the electronic width of $\rho$-meson. The QCD model
with infinite number of vector mesons is used. The more precisely determined
results are:  $a_{\mu}(hadr.) = 678(7).10^{-10}$, $\delta
\alpha_{hadr.}(M^2_z) = 0.02786 (6)$.

\end{abstract}

\bigskip

The purpose of this work is to determine more precisely the
contribution of strong interactions to the value of
$a_{\mu}(hadr.)$ and $\delta \alpha_{hadr.}(M^2_z)$ obtained in
papers [1,2]. There are two reasons to write this letter.

1. A new more exact value of the electronic width of $\rho$-meson appeared
[3]:

\be
\Gamma^{ee}_0 = 6.85 \pm 0.11 keV
\ee
instead of the value $\Gamma^{ee}_0 = 6.77 \pm 0.32 keV$ used in [1] and
of the value $\Gamma^{ee}_0 = 6.72 \pm 10 keV$ used in [2], obtained by the
author in paper [4], from the analysis of the old experiments on the
measurement of the pion electromagnetic form factor.

2. Function $R(s)$ is calculated by new formula obtained in paper [5] where
analiticity requirement of the QCD polarization operators were combined with
the renormalization group. Function $R(s)$ for 3 flavours has the form

\be
R(s) = \frac{3}{2} (1 + r(s)) \ee Function $r(s)$ is calculated
[5] using the renormgroup and the requirement of absence of
nonphysical singularities. All the formulae for calculation of the
function $r(s)$ are given in paper [5]. The QCD model with and
infinite number of vector mesons developed in papers [4,6,7] is
the basis for the calculations. This model allows one to satisfy
the requirements of the Wilson operator expansion and the
nonperturbative effects [8] are included into it automatically.
Such a model is very useful for calculation of the integrals of
the function

\be
R(s) = \frac{\sigma(e^+e^- \to hadrons)}{\sigma(e^+e^- \to \mu^+ \mu^-)}
\ee
Function $R(s)$ in this model has the form

\be
R(s) = R^{I=1}_{u d} (s) + R^{I=0}_{u d}(s) + R_s(s) + R_c(s) +
R_b(s) \ee Here $R^{I=1}_{u d}$ and $R^{I=0}_{u d}$ are the
contributions of $u$ and $d$ quarks in the state with isotopic
spin $I=1$ ($\rho$ family) and $I=0$ ($\omega$ family) and
$R_s(s)$, $R_c(s)$, $R_b(s)$ are the contributions of $s$
($\varphi$ family), $c$ ($J/\psi$ family) and $b$ ($\Upsilon$
family) quarks, respectively. The narrow resonancees approximation
is used. If the total width of the resonances $\Gamma_k \ll M_k$
($M_k$  is the k-th resonance mass) for all $k$, the narrow
resonances approximation is valid. If $M_k \Gamma_k \gg M^2_k -
M^2_{k-1}$, then beginning from the k-th resonance function $R(s)$
will be described by a smooth curve and all the formulae of the
model will be valid [9]. To calculate the integral of $R(s)$ in
the QCD model with infinite number of vector mesons one should
only know a few number of masses and electronic widths of lowlying
resonances in each family [1,2]. Comparing with [1,2] only
$R^{I=1}_{u d}(s)$ had been corrected. The changes  in the
contributions of the rest families are very small. All the
formulae of papers [1,2] except for the formulae for $R^{I=1}_{u
d} (s)$ are unchanged.

The results of the calculations for hadronic contribution of the muon
$(g-2)$ factor are the following:

\be
a_{\mu} (hadr.) = \frac{\alpha^2}{3 \pi^2}
\int\limits^{\infty}_{4m^2_{\pi}}~ds K(s) R(s)/s = 678
(7).10^{-10} \ee

where
$$
K(s) = x^2(1 - x^2/2) + (1 + x^2)(1 + x^{-2}) \Biggl [ln(1+x) - x+x^2/2
\Biggr ] +
$$
\be
\frac{1 + x}{1 - x} x^2 ln x; ~~ x = \frac{1 - (1 -
4m^2_{\mu}/s)^{1/2}}{1 + (1 - 4m^2_{\mu}/s)^{1/2}} \ee

The result (5) should be compared with the recent, the most exact
calculations $\alpha_{\mu}(hadr.)$ obtained by integrating formula
(5) with the experimental cross section of $e^+e^-$ annihilation
into hadrons [10,11].
\be
a_{\mu} (hadr.) = (684.7 \pm 7)\cdot 10^{-10} ~~~~ [10]
 \ee
\be
a_{\mu} (hadr.) = (681.1 \pm 6.3)\cdot 10^{-10} ~~~ [11]
 \ee
 For
hadronic contribution into $\alpha(M_z)$  we have obtained

\be
\delta \alpha_{hadr.} = \frac{\alpha M^2_z}{3 \pi} ~P
\int\limits^{\infty}_{4 m^2_{\pi}}~ \frac{R(s) ds}{(M^2_z - s)s} =
0.02786 (6) \ee

The result (8) can be compared with the results of papers [12-22]:

$$ \delta \alpha_{hadr.} = 0.02744 (36) ~~~~ [12] $$ $$ \delta
\alpha_{hadr.} = 0.02803(65) ~~~~ [13] $$ $$ \delta \alpha_{hadr.}
= 0.02780 (6)  ~~~~ [2] $$ $$ \delta \alpha_{hadr.} = 0.0280 (7)
~~~~~~ [14] $$ $$ \delta \alpha_{hadr.} = 0.02754 (46) ~~~~ [15]
$$ $$ \delta \alpha_{hadr.} = 0.02784 (22) ~~~~ [16] $$ $$ \delta
\alpha_{hadr.} = 0.02778 (16) ~~~~ [17] $$ $$ \delta
\alpha_{hadr.} = 0.02779 (20) ~~~~ [18] $$ $$ \delta
\alpha_{hadr.} = 0.02770 (15) ~~~~ [19] $$ $$ \delta
\alpha_{hadr.} = 0.02787 (32) ~~~~ [20] $$ $$ \delta
\alpha_{hadr.} = 0.02778 (24) ~~~~ [21] $$ $$ \delta
\alpha_{hadr.} = 0.02741 (19) ~~~~[22] $$ obtained by integrating
over eq.(8) with experimental cross section for $e^+e^-$
annihilation into hadrons.

\vspace{1cm}

\begin{center}
{\bf Acknowledgements}
\end{center}

\vspace{3mm} It was possible for us to do the research described
in this publication in part by the Award No.RP2-2247 of the US
Civilian Research  and Development Foundation for the Independent
States of the Former Soviet Union (CRDF), by the Russian Found of
Basic Research, Grant No.00-02-17808 and INTAS Call 2000, Project
No.587.


\newpage

\begin{thebibliography}{99}
\bibitem{1} B.V.Geshkenbein, V.L.Morgunov, Phys.Lett. {\bf B340} (1994) 185.
\bibitem{2} B.V.Geshkenbein, V.L.Morgunov, Phys.Lett. {\bf B352} (1995) 456.
\bibitem{3} Particle Data Group, K.Hagiwara et al., Phys.Rev. {\bf D66}
(2000) 010001-1.
\bibitem{4} B.V.Geshkenbein, Yad.Fiz {\bf 59} (1996) 309.
\bibitem{5} B.V.Geshkenbein, hep-ph/0206094. Phys.Rev. {\bf D67},
0540XX (2003).
\bibitem{6} B.V.Geshkenbein, Yad.Fiz. {\bf 51} (1990) 1121.
\bibitem{7} B.V.Geshkenbein, V.L.Morgunov, Yad.Fiz. {\bf 58} (1995) 1873.
\bibitem{8} M.A.Shifman, A.I.Vainstein and V.I.Zakharov, Nucl.Phys. {\bf
B147} (1979) 385.
\bibitem{9} B.V.Geshkenbein, Yad.Fiz. {\bf 49} (1989) 1138.
\bibitem{10} M.Davier, S.Eidelman, A.H\"ocker and L.Zhang, hep-ph/0208177.
\bibitem{ab}  K.Hagiwara, A.D.Martin, Daisuke Nomura and T.Teubner,
hep-ph/02091187~v2.
\bibitem{11} A.D.Martin, D.Zeppenfeld, Phys.Lett. {\bf B345} (1995) 558.
\bibitem{12} S.Eidelman, F.Jegerlehner, Z.Phys. {\bf C67} (1995) 585.
\bibitem{13} H.Burkhardt, B.Pietrzyk, Phys.Lett. {\bf B356} (1995) 398.
\bibitem{14} M.L.Swartz, Phys.Rev. {\bf D53} (1996) 5268.
\bibitem{15} M.Davier, A.H\"ocker, Phys.Lett. {\bf B419} (1998) 419.
\bibitem{16} J.H.K\"uhn, M.Steinhauser, Phys.Lett. {\bf B437} (1998) 425.
\bibitem{17} J.Erler, Phys.Rev. {\bf D59} (1999) 054008.
\bibitem{18} M.Davier, A.H\"ocker, Phys.Lett. {\bf B435} (1998) 427.
\bibitem{19} S.Groote et al., Phys.Lett. {\bf B440} (1998) 375.
\bibitem{20} F.Jegerlehner, hep-ph/9901386.
\bibitem{21} A.D.Martin, J.Outhwaite, M.G.Ryskin, Phys.Lett. {\bf B492}
(2000) 69.

\end{thebibliography}
\end{document}